\documentstyle[12pt]{article}
\begin{document} 
\begin{center}
{\bf \large 

Hierarchic trees with branching number close to one: noiseless KPZ
 equation with additional linear term for imitation of 2-d and 3-d phase transitions.}\\

\vspace{4mm}
D.B. Saakian\\

Yerevan Physics Institute,
Alikhanian Brothers St. 2,\\ Yerevan 375036, Armenia 

\end{center}

\begin{abstract}

An imitation of 2d field theory is formulated
by means of a model on the hierarhic tree (with branching number 
close to one) with the same 
potential and the free correlators identical to 2d correlators ones.
 Such a model carries on some features of the 
original model for certain scale invariant theories. For the case of 2d conformal
models it is possible to derive exact results. 
The renormalization group equation for the free energy is noiseless KPZ  equation 
with additional linear term.

\end{abstract}

One of the most fruitful ideas in physics is the idea of universality. 
In fact it is the only hope due to which rather artificial models
of the present theoretical physics can successfully capture the
relevant aspects of the nature.
We believe, that at the critical point statistical
mechanics system omits the secondary details. 
Usually this concerns the Hamiltonian in the $d$ 
dimensional Euclidean space. All Hamiltonians from the universality
class have the same multi point (2-point, three-point,4-point..) correlators for fluctuating fields.
These variables, as well as the phase structure have deep and universal nature.

We propose to follow another path: by keeping Hamiltonian fixed to simplify
 the space geometry as much as possible retaining  two point correlators and three point (for isiosceles triangles)correlators.
\\ 
If the action of original theory   consists of the Laplacian and a
potential, our model feels the space dimension through the behavior of Green
function
\begin{eqnarray}
\label{e1}
G(x,x')\sim \frac{1}{r(x,x')^{d-2}} d\ne 2\nonumber\\
G(x,x')\sim \log\frac{1}{r(x,x')}, d=2
\end{eqnarray}
The total volume is 
\begin{equation}
\label{e2}
(\frac{L}{a})^d
\end{equation}
where {\it L} and {\it a} are the infrared and ultraviolet cutoffs, $r(x,x')$ is the distance.
The Euclidean geometry contains too much constructions. One can rotate a point around 
some center and write out close circle. 
Let us now consider some metric space with properties: \\
A.For every pair of points there is a distance r(x,x').\\
B.We have some measure at every point $d\mu_s(x)$ with the total measure $\int d \mu_s=R^d$.\\
C.One can construct a quadratic form with corresponding asymptotics (1) for Green function.\\
We are going to construct statistical mechanics models on the 
simplest space, which supports points A-C.
We hope, that due to the universality these models will 
acquire some properties of models in d-dimensional space. 
To realize this program we will use certain ideas from the theory
of Random Energy Model (REM) [1-5].
In ref. [5]  a relation of 2d quantum Liouville model to REM 
and to the Directed Polymer (DP) on Cayley tree was established \\
Our present analysis shows, that the connection with REM is not a 
specific feature of Liouville model and works well also for other 
conformal models. Moreover, 
using similar ideas we intend to 
construct general 2-d quantum models in the ultrametric space and thereby generalize the 
above-mentioned 
connection between the quantum field theoretical models and 
those defined on the hierarchical lattices.\\
Let us consider a hierarchic tree with the branching number q. We begin with integer q, then 
continue the obtained expressions analytically to the point $q\to 1$. Instead of d-d Euclidean space now we have 
$q^K$ endpoints, where K is a number of hierarchic levels.
First we define the fields $f_l$ on branches of a tree. The field $\phi$ at the endpoint {\it x}
is defined as 
\begin{equation}
\label{e3}
\phi(x)=f_0+\sum_lf_l
\end{equation}
 The 
summation in (3) is made
along the trajectory of point x, connecting it with the origin of the tree. 
We define v at the hierarchy level j as 
\begin{equation}
\label{e4}
v=\frac{jV}{K}
\end{equation}
Now determine the kinematic part of the action for the field $\phi(x)$ 
\begin{equation}
\label{e5}
\sum \frac{K}{V}f(v,l)^2
\end{equation}
Then the partition under the potential $U(\phi)$ is
\begin{eqnarray}
\label{e6}
\int {\it d}f\exp\{-\sum_{v=0}^{V}\frac{K}{2V}
f(v,l)^2\}\nonumber\\
\exp\{\sum_xU(\phi(x))\}
\end{eqnarray}
We have for the correlator 
\begin{equation}
\label{e7}
<\phi(x)\phi(x')>=v
\end{equation}
For usual 2d models with 
\begin{equation}
\label{e8}
\int {\it d}\phi_0{\it d}\phi\exp\{-\frac{1}{8\pi} {\it d}x^2\nabla \phi(x)^2\}
\exp\{\int d x U(\phi(x))\}
\end{equation}
the total surface area is equal to $R^2$, and the correlators read as
\begin{equation}
\label{e9}
<\phi(x)\phi(x')>=\ln \frac{L^2}{r^2}
\end{equation} 
It is possible to take n component fields in Eq. (8) instead of the one-component field $\phi(x)$.\\
We can determine the distance from the equality $V=\ln\ r^2$. Then our
correlators coincide.
What is the advantage of 
representation (6)? We are in a position to calculate the partition function
by means of iterations. This is well known for models on hierarchical
lattices [6]. 
Let us take some large number K and derive 
\begin{eqnarray}
\label{e10}
I_1(x)=\sqrt{\frac{K}{2V\pi}}\int _{-\infty}^{\infty}\exp\{-\frac{K}{2V} y^2+U(x+y)\}{\it d }y\nonumber\\
I_{i+1}(x)=\sqrt{\frac{K}{2V\pi}}\int _{-\infty}^{\infty}\exp\{-\frac{K}{2V} y^2\}
[I_{i}(x+y)]^{q}{\it d }y\nonumber\\
Z=\lim_{K\to \infty}[I_{K}(0)]^{q}
\end{eqnarray}
As for the determination of partition function, 
we need only the equation (10) and  can define our model for any value of q consideration
of the 
analytical continuation of (10).
Let us consider the limit 
\begin{eqnarray}
\label{e11}
q\to 1, K\to\infty\nonumber\\
q^K=e^V
\end{eqnarray}
Using the small factor $(q-1)$, it is possible to introduce continuous measures $d\mu _x, d \mu_l$  
construct construct perturbative 
field theory on this ultrametric space, and  calculate diagrams. In reality we need in expressions for the propogator (7), as well as the total volume measure 
inside the sphere with maximal hierarchic distance v given by equality
\begin{eqnarray}
\label{e12}
\int d\mu_x=e^v-1
\end{eqnarray}
For finite or large values of q considered in [5] it is impossible (or too difficult) to 
define the perturbative regime.

Let us consider  carefully equation (10) at the limit (11).
We introduce a variable $w(v,x)=I_{\frac{Kv}{V}}(x)$ and consider the limit $\frac{V}{K}\ll 1$. For the differential $dv$ we have the expression $\frac{V}{K}$. Let us take also 
\begin{eqnarray}
\label{e13}
q-1=\frac{V}{K}\equiv dv
\end{eqnarray}
Using expression $x^q\approx x(1+\log x(q-1))$ it is easy to obtain
\begin{eqnarray}
\label{e14}
\frac{d w}{d v}=w\ln w +\frac{1}{2}\Delta w
\end{eqnarray}
After the replacement $w=\exp(u(t,x)$ we arrive at
\begin{eqnarray}
\label{e15}
\frac{d u}{d v}=\frac{1}{2}\Delta u+\frac{1}{2}(\nabla u)^2+u\nonumber\\
u(0,x)=U(x)
\end{eqnarray}
where $U(x)$ is a potential in Eq.(8). The dimension n of the space where this equation is formulated is equal to the number of different fields $\phi(x)$ in Eq. (8). Having an expression for $u(v,x)$ we obtain 
for the free energy 
\begin{eqnarray}
\label{e16}
\ln Z=u(V,0)
\end{eqnarray}
For the free energy $u(v,x)$ we have the noiseless KPZ equation (15) with an additional linear term .\\
There are two interesting solution of Eq. (15) at large values v. If the couplings in the polynomial potential are $O(1)$,it is reasonable to consider the solution at large values of v (and far from the renormalization group fixed points):
\begin{equation}
\label{e17}
u(v,x)=const \exp(v)
\end{equation}
If one considers the couplings $\sim \frac{1}{\exp(V)}$ in the potential $U(x)$ , then 
a solution
  a solution:
\begin{equation}
\label{e18}
u(v,x)=const\exp(v)+u_s(x), u_s(x)\sim 1.
\end{equation} 
\begin{equation}
\label{e19}
\frac{1}{2}\Delta u_s+\frac{1}{2}(\nabla u_s)^2+u_s=0
\end{equation}
corresponds to the perturbative regime
This equation gives the efficient potential at the stable point of renormalization group.
One can rewrite Eq. (19) in another form for the $z\equiv \frac{d u_s}{dv}$:
\begin{equation}
\label{e20}
\frac{dz}{du_s}+z+\frac{2}{z}u_s
\end{equation}
In analogy to Eqs. (10),(14) it is also possible to derive the correlators. To calculate the correlator 
$<\exp(i\alpha\phi(x)-\alpha\phi(y))>$ , where the hierarchic distance between points $x,y$
is $v_0$, one has to consider also an equation
\begin{eqnarray}
\label{e21}
\frac{d f(v,x,\alpha)}{d v}=f\ln w +\frac{1}{2}\Delta f\nonumber\\
f(0,x,\alpha)=\exp(U(x)+i\alpha x)
\end{eqnarray}
Then for the generating function $f_0(v,x)$ of correlator one should solve again Eq. (21) with the boundary conditions at the point $v_0$
\begin{eqnarray}
\label{e22}
f_0(v_0,x)=f(v_0,x,\alpha)f(v_0,x,-\alpha)/w(v_0,x)
\end{eqnarray}
For the correlator we obtain an expression:
\begin{eqnarray}
\label{e23}
<\exp(i\alpha\phi(x)-\alpha\phi(y))>=\frac{f_0(\infty,0)}{w(\infty,0)}
\end{eqnarray} 
Let us use the same approach for the case of $d>2$.
For the volume in d-d space one has $\sim a^d$. If we identify it with our $q^L$, then derive
$a=q^{\frac{1}{d}L}$. 
The fields $f_l$ are defined on the branches of tree, $f_0,f_1$ at the origin.
Let us define the free field action. 
\begin{equation}
\label{e24}
\phi(x)=f_0+f_1+\sum_{vl} f(v,l)
\end{equation}
The summation  in (24) is along the trajectory of point X. 
Now determine the kinematic part of the action for the field $\phi(x)$ 
\begin{equation}
\label{e25}
A=\frac{1}{2}[{f_1}^2+ \sum_{vl} \exp(-\alpha v) f(v,l)^2/\alpha]
\end{equation}
 If one takes 
$\alpha=\frac{d-2}{d}$ for the combined field,
then
\begin{eqnarray}
\label{e26}
<\phi(x)\phi(x')>=\exp(\alpha v) \sim \frac{L}{r(x,x')}^{(d-2)},
\end{eqnarray}
where L is the maximal distance in the model (the infrared cutoff).
Now (10) transforms into:
\begin{eqnarray}
\label{e27}
I_1(x)=\sqrt{\frac{K}{2V\alpha \pi}}\int _{-\infty}^{\infty}\exp\{-\frac{K}{2V\alpha} y^2+U(x+y)\}{\it d }y\nonumber\\
I_{i+1}(x)=\sqrt{\frac{K}{2V\alpha e^{\alpha V(K-i+1)/K}\pi}}\int _{-\infty}^{\infty}e^{\{
-\frac{K}{2V\alpha \exp[\alpha V(K-i+1)/K]} y^2\}}
[I_{i}(x+y)]^q{\it d }y\nonumber\\
Z=\lim_{K\to \infty}\sqrt{\frac{1}{2\pi}}\int _{-\infty}^{\infty}\exp\{-\frac{1}{2} y^2+U(y)\}{\it d }y[I_{K}(y)]^{q}
\end{eqnarray}
To calculate $I_K(x)\equiv w(V,x)\equiv \exp(u(V,x))$ we should solve the equation like (15)
\begin{eqnarray}
\label{e28}
\frac{d u}{d v}=\frac{1}{2}\alpha \exp[\alpha v]\Delta u+\frac{1}{2}(\nabla u)^2+u\nonumber\\
u(0,x)=U(x)
\end{eqnarray}
We gave a simplified, approximate method for the 2d field theoretical models. We hope, that the bulk structure, the two and three point correlators (for isosceles
 triangles) are the same, as in 2d models. We checked, that three point correlators are the same, as those in 2d case. According to [5], the model with $U(x)=\exp (kx)$ on such hierarchic lattices at large values of {\it q} exactly is equivalent to Liouvi
lle model. According to results of [8] the thermodynamic limit of this model is independent of q. Thus at least for these case models on our tree (branching number $q$ is close to one) are equivalent to those in 2-d.
 It is  possible to check our hypothesis about 
 equivalence of  models on our trees with some segment of dd field theory by means of 
 direct numerical calculation of Eq. (15),(28), for example for field version of 3d Ising model 
 with proper choice of potential U. \\
I am grateful to  ISTC grant A-102 
 for partial financial support, C. Lang and W. Janke for invitation to Graz and Leipzig
and discussions, P. Grassberger for useful remark. 

\end{document}